\documentclass[sigconf]{acmart}

\usepackage{siunitx} 

\usepackage[linesnumbered,ruled,vlined]{algorithm2e}
\usepackage[noend]{algorithmic}

\usepackage{multirow}
\usepackage{bm}
\usepackage{stfloats}
\usepackage{hyperref}

\AtBeginDocument{%
  }

\copyrightyear{2024}
\acmYear{2024}
\setcopyright{acmlicensed}
\acmConference[ICAIF '24]{5th ACM International Conference on AI in Finance}{November 14--17, 2024}{Brooklyn, NY, USA}
\acmBooktitle{5th ACM International Conference on AI in Finance (ICAIF '24), November 14--17, 2024, Brooklyn, NY, USA}
\acmDOI{10.1145/3677052.3698683}
\acmISBN{979-8-4007-1081-0/24/11}

\begin{document}

\title{Rolling Forward: Enhancing LightGCN with Causal Graph Convolution for Credit Bond Recommendation}

\author{Ashraf Ghiye}
\orcid{0009-0008-7088-0337}
\affiliation{%
  \institution{BNP Paribas CIB}
  \department{Global Markets, Data \& AI Lab}
  \city{Paris}
  \country{France}
}

\affiliation{%
  \institution{École Polytechnique}
  \department{Computer Science Laboratory, LIX}
  \city{Palaiseau}
  \country{France}
}


\author{Baptiste Barreau}
\orcid{0000-0001-9045-0141}
\affiliation{%
  \institution{BNP Paribas CIB}
  \department{Global Markets, Data \& AI Lab}
  \city{Paris}
  \country{France}
}

\author{Laurent Carlier}
\orcid{0009-0000-0631-6100}
\affiliation{%
  \institution{BNP Paribas CIB}
  \department{Global Markets, Data \& AI Lab}
  \city{Paris}
  \country{France}
}

\author{Michalis Vazirgiannis}
\orcid{0000-0001-5923-4440}
\affiliation{%
  \institution{École Polytechnique}
  \department{Computer Science Laboratory, LIX}
  \city{Palaiseau}
  \country{France}
}

\authorsaddresses{Correspondence:
\href{mailto:ashraf.ghiye@bnpparibas.com}{ashraf.ghiye@bnpparibas.com}}

\begin{abstract}
Graph Neural Networks have significantly advanced research in recommender systems over the past few years. These methods typically capture global interests using aggregated past interactions and rely on static embeddings of users and items over extended periods of time. While effective in some domains, these methods fall short in many real-world scenarios, especially in finance, where user interests and item popularity evolve rapidly over time. To address these challenges, we introduce a novel extension to Light Graph Convolutional Network (LightGCN) designed to learn temporal node embeddings that capture dynamic interests. Our approach employs causal convolution to maintain a forward-looking model architecture. By preserving the chronological order of user-item interactions and introducing a dynamic update mechanism for embeddings through a sliding window, the proposed model generates well-timed and contextually relevant recommendations. Extensive experiments on a real-world dataset from BNP Paribas demonstrate that our approach significantly enhances the performance of LightGCN while maintaining the simplicity and efficiency of its architecture. Our findings provide new insights into designing graph-based recommender systems in time-sensitive applications, particularly for financial product recommendations.
\end{abstract}

\ccsdesc[500]{Information systems~Recommender systems}
\ccsdesc[500]{Information systems~Collaborative filtering}

%
\keywords{Graph Neural Networks, Dynamic Recommendation, Credit Bond, Recommender Systems, Finance, Collaborative Filtering}

\maketitle
\renewcommand{\shortauthors}{A. Ghiye et al.}

\section{Introduction}

Graph Neural Networks (GNNs) have emerged as the state-of-the-art for recommender systems due to their ability to model complex interactions in user-item networks~\cite{Anelli2023, Gao2023, Wu2022, Ying2018}. Unlike traditional collaborative filtering techniques like Matrix Factorization~\cite{Koren2009}, which fail to capture high-order signals~\cite{Yang2018}, GNNs leverage structural information to learn enriched node representations. By using edges to facilitate the propagation, aggregation and update of these representations, GNNs can effectively harness the collaborative signal explicitly embedded in the graph structure~\cite{Wang2019, He2020}. 

Despite these advancements, graph recommender systems largely overlook the impact of time and the order of interactions—key factors in the design, training, and evaluation of dynamic recommender systems. Most existing GNN models fail to account for dynamic changes in the graph structure~\cite{Gao2023}, which limits their effectiveness in time-sensitive settings where user interest and item popularity evolve quickly. This limitation is especially important in finance. For instance, Corporate and Institutional Banks need recommendation systems to provide their clients with relevant time-aware advisory services. Thus, these systems must swiftly adapt to clients' changing needs and fluctuating item popularity to help salespeople deliver tailored recommendations to their clients. 

To address these challenges, we propose a novel framework to train Graph Convolutional Networks~\cite{Kipf2017} for recommendation. Our approach uses a discrete-time representation of the dynamic graph, with each snapshot serving as a temporal batch. Instead of utilizing the entire graph during training, we employ causal convolutions to capture present user preferences from past snapshots. Additionally, we fix a predetermined window size to control the effect of data drift and maintain the quality of recommendations.

We implement our method using LightGCN, a simple yet powerful framework for Graph Collaborative Filtering~\cite{He2020}. Our approach achieves significant performance gains, improving mean Average Precision by up to 4x over the conventional training method, without compromising the simplicity and lightweight nature of LightGCN. Through comprehensive experiments on a real-world credit bonds dataset, we examine the effect of the window size on model performance and validate the effectiveness of our method in delivering relevant recommendations in financial applications.


%
%

\section{Background and Related Work}

\subsection{Background}
The primary goal of a recommender system is to estimate the likelihood of a user $u$ showing interest towards a particular item $i$. The system should be capable of scoring any user-item pair from the set of all users $U$ and all items $I$ at any given time $t$. Historical preferences, denoted by the triplets $(u,i,t)$, form a three-dimensional tensor $A$ such that $a_{ui}^t$ represents the feedback provided by a user $u$ to an item $i$ at time $t$. Often, the temporal dimension is collapsed, resulting in a static matrix of interactions $A$. In our study, users correspond to institutional clients and items to credit bonds. The feedback represents various historical indications of interest.

Collaborative Filtering (CF), a prominent technique in recommender systems, assumes that two users who have same preferences towards certain items will likely have similar preferences towards other ones~\cite{Goldberg1992}. Many algorithms have been introduced over the years to formalize and develop this idea further. One of the most notable algorithms is Matrix Factorization (MF). The core idea of MF is to learn latent representations for each user $\pmb{e}_u \in \mathbb{R}^d$ and item $\pmb{e}_i \in \mathbb{R}^d$, such that their dot product approximates the interaction matrix ($\hat{a}_{ui} =\pmb{e}_u \cdot \pmb{e}_i \approx a_{ui}$)~\cite{Koren2009}. In MF, these latent representations, also known as ID embeddings, are shallow encodings that represent user and item IDs as vectors, stored in look-up tables: $\pmb{e}_u = f(u), \, \pmb{e}_i = f(i)$. Once optimized, these latent representations are assumed to capture semantic meanings, such as genres in the context of movie recommendations.

With the rise of deep learning, a second generation of collaborative filtering has emerged. Neural Collaborative Filtering (NCF) replaces the encoding function $f$ with neural networks, allowing models to capture more complex non-linear patterns in the data and incorporate side information like content features~\cite{He2017}. A third generational improvement consists of employing Graph Neural Networks (GNNs) to enrich the embeddings using the structure of the interaction matrix $A$~\cite{Wang2019}. Today, Graph Collaborative Filtering methods are considered the state-of-the-art as they explicitly encode the collaborative signals in the learning process~\cite{Wang2019, He2020, Wu2022}.

\subsection{Related Work}
Despite these advancements, time remains a crucial yet largely overlooked factor in collaborative filtering techniques~\cite{Borba2017, Campos2014}. This oversight leads to an oversimplification of user preference modeling and potential data leakage~\cite{Sun2023, Ji2023}. According to~\cite{Sun2023}, only a few studies maintain the chronological order of interactions and consider the absolute time points in their predictions. Moreover, modeling interactions occurring years apart similarly to those occurring closer in time contradicts the growing body of evidence suggesting that user preferences are more likely to be similar within shorter time frames~\cite{Ding2005, Nzeko2017}. While many studies have explored ways to integrate time into traditional~\cite{Ding2005, Koren2009b, Xiong2010} and neural-based~\cite{Ji2019, Liu2020} approaches, fewer have addressed this issue in graph-based approaches~\cite{Li2020}. 

Alternatively, other studies have critically examined the reliance on extensive historical data, proposing strategies to cope with data drift in dynamic environments~\cite{Bogina2023} by either disregarding older data~\cite{Verachtert2022, Barreau2020, Nasraoui2007} or applying fading factors~\cite{Ding2005, Nzeko2017, Ghiye2023}. For example, \cite{Verachtert2022} showed that using only recent interactions can drastically enhance the performance of traditional recommender systems, particularly for online news. In financial recommendations, \cite{Barreau2020} used sampled sets of historical interactions to build context-aware user profiles, and \cite{Ghiye2023} extended this approach by applying learnable factors to maintain the relevance and accuracy of recommendations over time. However, their methods only capture first-order signals.

Recent developments in temporal graph learning hold promise for improving the modeling of dynamic graphs. These models generally rely on complex time modules to learn dynamic node embeddings. Despite their potential, they are primarily designed for link prediction tasks~\cite{Rossi2020, Xu2020, You2022}, and their application to recommender systems and ranking tasks remains largely underexplored; to the best of our knowledge, \cite{Kim2024} is the first study to apply a Temporal Graph Network (TGN) for personalized ranking. Additionally, our evaluation process necessitates exhaustive scoring of user-item pairs over time, making evaluations with complex architectures like TGN computationally infeasible. Studies on dynamic link predictions often simplify their evaluation protocols by considering only a small fraction of negative pairs, failing to capture the full dimension and complexity of the problem. Subsequent research has highlighted the pitfalls of current evaluation protocols used in dynamic link prediction literature, proposing more robust protocols that could significantly alter model performance, raising concerns about the actual added value of their complex components~\cite{Poursafaei2022, Daniluk2023}. 


Our study focuses on extending LightGCN, originally developed for static graphs, to dynamic applications in the financial sector. By introducing a novel causal convolutional approach over temporal snapshots, we aim to integrate temporal modeling effectively into graph-based collaborative filtering and demonstrate its practical applicability in providing time-aware financial recommendations.

\section{Methodology}

\begin{figure*}[!h]
\centering
\includegraphics[width=1.2\columnwidth]{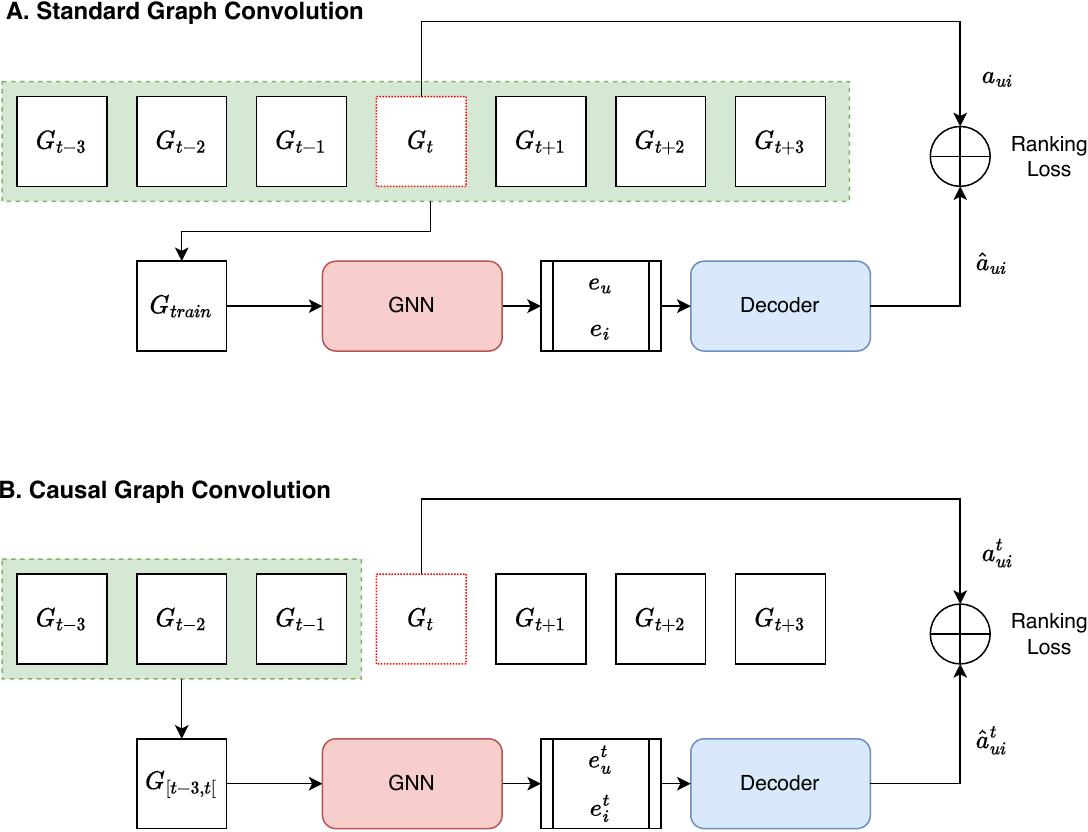}
\caption{\small Standard training framework (A), where the full graph is used to predict each batch of events; this approach learns one static embedding per node in the graph and uses these embeddings to learn a static ranking over the full period. Our approach (B) consists of using a rolling window to learn dynamic ranking; the model has to be always forward-looking, and the node embeddings change from one period to another, allowing them to capture the dynamic context.}
\label{fig:model_schema}
\end{figure*}

Our method extends the common training approach used by static GNNs (Fig.\ref{fig:model_schema}.A), ensuring the model learns from historical interactions while avoiding potential biases from considering future ones during training. To that end, we propose constructing the training graph dynamically using only recent transactions within $[t-w,t[$ to predict those occurring at $t$. Our approach (Fig.\ref{fig:model_schema}.B) preserves the integrity and order of the interactions, aligning the training process with how the model serves recommendations in reality, that is, by incorporating incoming data incrementally and limiting access to previously seen interactions.

Formally, let $\mathcal{E}_{t} = \{\,(u,i,t’) \mid t’ = t\,\}$ be the set of transactions on day $t$, where $u$ and $i$ are used to denote a user and an item, respectively. Each transaction set is associated with a bipartite graph, $G_t(V_t, \mathcal{E}_t)$, where $V_t$, the set of nodes, is the union of two disjoint sets: users $U_t$ and items $I_t$. The user set is considered relatively stable over time, while the item set changes more frequently—items can be added or deleted from one period to another. 


\subsection{Standard Graph Convolution}
It has been the convention to split the dataset based on a single time cut\footnote{Alternative data splitting schemes include random splitting, where a percentage of each user's transactions are randomly sampled as test instances, and leave-one-out approach where the last transaction(s) of each user serve as the test instance.} into two disjoint sets: $\mathcal{E}_{train} = \{\,(u,i, t') \mid t' \leq t_{cut}\,\}$, and $\mathcal{E}_{test}=\{\,(u,i,t') \mid t'>t_{cut}\,\}$, where each set is associated with its own graph, $G_{train}$ and $G_{test}$. Consequently, a graph-based algorithm uses the full graph $G_{train}$ to learn and optimize the node embeddings, and the test graph $G_{test}$ to evaluate performance.

The following reasoning can be readily applied to any static GNN-based model; however, we will illustrate it using LightGCN due to its competitive performance and remarkable efficiency, making it highly scalable and suitable for real-time financial recommendations. Unlike other GNN models, LightGCN eliminates feature transformation and nonlinear activation, focusing on the graph convolution operation. Consequently, LightGCN learns enriched node representations by linearly propagating the embeddings on the training graph, $G_{train}$, to capture the collaborative signals encoded in its structure. The message-passing equations can be simply written as follows:

\begin{equation}
\begin{aligned}
    \pmb{h}_u^\text{(k)} = \sum_{i \in \mathcal{N}(u)} c_{ui} \; \pmb{h}_i^\text{(k-1)}, \\
    \pmb{h}_i^\text{(k)} = \sum_{u \in \mathcal{N}(i)} c_{ui} \; \pmb{h}_u^\text{(k-1)},
   \end{aligned}
\end{equation}

where $\pmb{h}^\text{k}_*$ denotes the hidden representation of node $*$ at layer $k$, $c_{ui}$ is a normalizing term that helps stabilize the embeddings' norm, and $\mathcal{N}(u) = \{\,j \mid (u,j) \in \mathcal{E}_{train}\,\}$ is the set of items connected to user $u$ in $G_{train}$, also referred to as the node neighborhood. $\mathcal{N}(i)$ is defined similarly. The representations at different layers capture varying semantics: the first layer aggregates information about direct interactions, the second layer aggregates information from users (items) that have common interests (consumption history), and so on. 

By default, the initial feature vector $\pmb{h}_*^\text{(0)} = x_* \in \mathbb{R}^d$ represents the node's ID embedding, where $d$ is the embedding size. If more features are used, the ID embedding and the feature vector(s) are concatenated and projected using a linear layer to maintain a unified embedding size for users and items. 

The final embedding of a user (item) node is taken as the weighted sum of its initial feature vector and all its $L$ hidden representations to capture different semantics and making the final embedding more comprehensible:

\begin{equation}
\label{eq:layer_aggregation}
    \pmb{e}_u = \sum_{k=0}^L \alpha_k \; \pmb{h}_u^\text{(k)}; \quad \pmb{e}_i = \sum_{k=0}^L \alpha_k \; \pmb{h}_i^\text{(k)},
\end{equation}


where $\alpha_k = \frac{1}{k+1}$. The model prediction, defined as the dot product of the final user and item representations, $\hat{a}_{ui} = \pmb{e}_u \cdot \pmb{e}_i$, is used as the ranking score to generate recommendations. 

\subsection{Causal Graph Convolution}
The previous approach has two main flaws: (1) each node has only one static embedding that encapsulates its activity over the full training period, and (2) information about future links is used to predict earlier ones in mini-batches, which might potentially cause data leakage. 

To address these issues, we extend this framework by introducing a temporal dimension to account for dynamic interests. Consequently, the score becomes $\hat{a}_{ui}^t = \pmb{e}_u^t \cdot \pmb{e}_i^t$. This allows the model to learn a dynamic ranking by learning two different temporal embeddings for the same user (item) at two different periods of time, as it is most likely that the user interest and/or the item popularity has shifted. 

To predict user interest at time $t$, a graph $G_{<t}$ is constructed such that only the transactions that occurred before $t$ are used to build its connectivity. In other words, the model is constrained to use a causal set of information when predicting her current interest. Hence, the message-passing equations are modified so that the graph $G_{<t}$, instead of $G_{train}$, is used when learning the temporal node representations at $t$:

\begin{equation}
\begin{aligned}
    \pmb{h}_u^\text{(k)}(t) = \sum_{(i, \, \Delta t) \in \mathcal{N}_{<t}(u)} c_{ui}^{\Delta t} \;  \pmb{h}_i^\text{(k-1)}(t), \\
    \pmb{h}_i^\text{(k)}(t) = \sum_{(u, \, \Delta t) \in \mathcal{N}_{<t}(i)} c_{ui}^{\Delta t} \;  \pmb{h}_u^\text{(k-1)}(t),
   \end{aligned}
\end{equation}

where $\mathcal{N}_{<t}(u) = \{\,(j, t-t') \mid (u,j,t') \in \mathcal{E}_{<t}\,\}$ is the set of causal neighborhood at time $t$, and $\Delta t = t - t' > 0$ is the relative time between prediction time and the time a transaction happened.

Finally, to control the effect of data drift and increase the model responsiveness to emerging interests and trends, the temporal graphs are constrained further by considering only the data lying within an interval of $w$ days from the prediction day. We define the temporally constrained sets of transactions at day $t$ as $\mathcal{E}_{t,w} = \{\,(u,i,t') \mid t' \in [t-w,t[\,\}$, and their associated graphs are now used to propagate and learn the node embeddings:

\begin{equation}
\begin{aligned}
    \pmb{h}_u^\text{(k)}(t) = \sum_{(i, \, \Delta t) \in \mathcal{N}_{t,w}(u)} c_{ui}^{\Delta t}  \; \pmb{h}_i^\text{(k-1)}(t), \\
    \pmb{h}_i^\text{(k)}(t) = \sum_{(u, \, \Delta t) \in \mathcal{N}_{t,w}(i)} c_{ui}^{\Delta t} \; \pmb{h}_u^\text{(k-1)}(t).
   \end{aligned}
\end{equation}

The window size, $w$, is a domain-specific hyper-parameter. Smaller values of $w$ might risk missing useful signals from older data, such as long-term preferences. Conversely, larger values might introduce noise or hinder the model's ability to capture short-term preferences, especially in domains with high data drift. In domains like news recommendation and finance, $w$ tends to be small due to the short-lived nature of information. However, in other domains like music and movie recommendation, $w$ can be larger as user tastes tend to shift more slowly.


For the following, we consider the problem of daily product recommendations, where $t$ refers to days. 

\section{Experimental Settings}
\label{section:model}

For the main models, we distinguish between three variants:
\begin{itemize}{
\item \textbf{LightGCN}: using the original architecture, we train the model using the standard training framework (Figure~\ref{fig:model_schema}.A). The normalizing term is set to $c_{ui} = \frac{1}{\sqrt{|\mathcal{N}(u)|} \sqrt{|\mathcal{N}(i)|}}$;
\item \textbf{LightGCN-W}: keeping the same architecture as above, but we train the model using our new approach (Figure~\ref{fig:model_schema}.B). The normalizing term is adapted to $c_{ui}^{\Delta t} = \frac{1}{\sqrt{|\mathcal{N}_{t,w}(u)|} \sqrt{|\mathcal{N}_{t,w}(i)|}}$;
\item \textbf{LightGCN-FW}: we modify the architecture by using the relative time of interactions to prioritize recent events, i.e., we apply time-aware coefficients of the form $c_{ui}^{\Delta t} = \frac{1}{\Delta t}$. 
}
\end{itemize}

Additionally, we use two standard benchmarks: a matrix factorization (\textbf{MF}) algorithm and a non-personalized popularity baseline, where items are ranked based on all the interactions in the training set (\textbf{MostPop}). We also include a more practical version of MostPop, where the popularity of an item is dynamically updated over time by considering its frequency in the last $w$ days (\textbf{RecentPop}~\cite{Ji2020}).

Unless otherwise specified, we use a single layer and only ID embeddings for GNN models. 

\subsection{Data}

We run our experiments on a proprietary dataset provided by BNP Paribas, comprising more than 7 million daily transactions spanning a period of five and a half years. The transactions correspond to Request for Quotations (RFQs) and Indication of Interests (IOIs) that clients show towards credit bonds. Overall, the dataset has over 5,000 unique clients and 47,000 unique bonds. Notably, the bond inventory is dynamic, with dozens of bonds being issued or maturing every day. Conversely, the client base remains relatively stable over time. Table~\ref{tab:dataset_summary} summarizes the average daily statistics. 

In addition to its ID, each bond is characterized by seven categorical features, namely its rating, sector, industry, country, currency, security grade, and seniority. 

Finally, the transactions are sorted chronologically and divided into three sets: (1) \textbf{Training set}: 01/01/2019 to 31/05/2022; (2) \textbf{Validation set}: 01/06/2022 to 31/12/2022; and (3) \textbf{Test set}: 01/01/2023 to 01/06/2023.

\begin{table}[!htbp]
\centering
\caption{\small Average statistics of the daily snapshots. 13.41\% of the transactions repeat from the previous day.}
\label{tab:dataset_summary}
\begin{tabular}{@{}lccccc@{}} 
\toprule
Dataset & $\bar{|\mathcal{E}_t|}$ & $|\bar{U_t}|$ & $|\bar{I_t}|$ & Period & Repeating \\
\midrule
Credit Bond & 6,226 & 771 & 4,425 & 1152d & 13.41\% \\
\bottomrule
\end{tabular}
\end{table}




\subsection{Training}
\label{section:training}

All models, except for MostPop and RecentPop, are trained using the Bayesian Personalized Ranking (BPR) loss~\cite{Rendle2009}, defined as:

\begin{equation}
    \label{eq:bpr_loss}
    \begin{split}
    \mathcal{L}_{BPR} = - \sum_{(t,u,i,j)\in \mathcal{D}_t(u)} \log(\sigma(\hat{a}_{ui}^t - \hat{a}_{uj}^t)),
    \end{split}
\end{equation}

where $\mathcal{D}_t(u)$ denotes the set of all possible quadruplets $(t,u,i,j)$, such that $(u,i,t) \in \mathcal{E}_t$ and $(u,j,t) \notin \mathcal{E}_t$. In other words, for every positive interaction in $\mathcal{E}_t$, we add all other valid but unobserved interactions to the set $\mathcal{D}_t(u)$. A valid negative interaction $(u,j,t)$ is one where the product $j$ has not reached maturity at time $t$ and not been interacted with by the user $u$ at that time.

The objective of this loss function is to learn a personalized ranking, ensuring that each user's positive interactions are scored higher than that of all their negative ones. The loss is commonly approximated with negative sampling. In our case, we use Dynamic Negative Sampling (DNS)~\cite{Zhang2013} with a 10:1 negative to positive ratio. 

We use the Adam optimizer~\cite{Adam2014} to train the models and employ early stopping to prevent overfitting. We set the embedding size to $64$ for both user and item IDs, and $16$ for each categorical feature when applicable. Additional implementation details and a pseudo-code are provided in Appendix~\ref{appendix:implementation_details} and Appendix~\ref{appendix:algorithm}, respectively.
  
\subsection{Evaluation}
\label{section:eval_metrics}

We address the challenge of optimal item ranking, commonly known as Top-k recommendation. Each day, our model generates a recommendation list for every user who has engaged in at least one transaction on that day, by ranking all the available items based on the output scores. The ranking quality of each list is then evaluated using four key metrics (formally defined in Appendix~\ref{appendix:detailed_metrics}):



\textbf{Mean Reciprocal Rank (MRR)~\cite{Buttcher2010}}: Measures the reciprocal of the rank of the first relevant item in the recommendation list.

\textbf{Recall@K~\cite{Shani2011}}: Calculates the proportion of relevant items that are included in the top-K recommended items, representing the model's ability to retrieve relevant items within the top-K.

\textbf{Mean Average Preicions (mAP)~\cite{Buttcher2010}}: Provides a holistic measure of precision at different recall levels. It measures the average precision of the recommendation list, taking into account the ranking of all relevant items. 

\textbf{Normalized Discounted Cumulative Gain (NDCG@K)~\cite{Shani2011} }: Evaluates the ranking quality of the top-K ranked items, discounted by the rank at which they appear. 


All these metrics range from 0 to 1, with 1 indicating a perfect ranking where all relevant items are placed at the top of the list. For NDCG and Recall, we choose $k=50$. Finally, the evaluation is repeated daily throughout the test period, and the final performance metrics are reported as the temporal average of these daily metrics, providing a comprehensive measure of performance over time. 

\section{Results}

%

\subsection{Window Size Analysis}
\label{section:history_size}

\begin{figure}[!h]
\centering
\includegraphics[width=0.92\columnwidth]{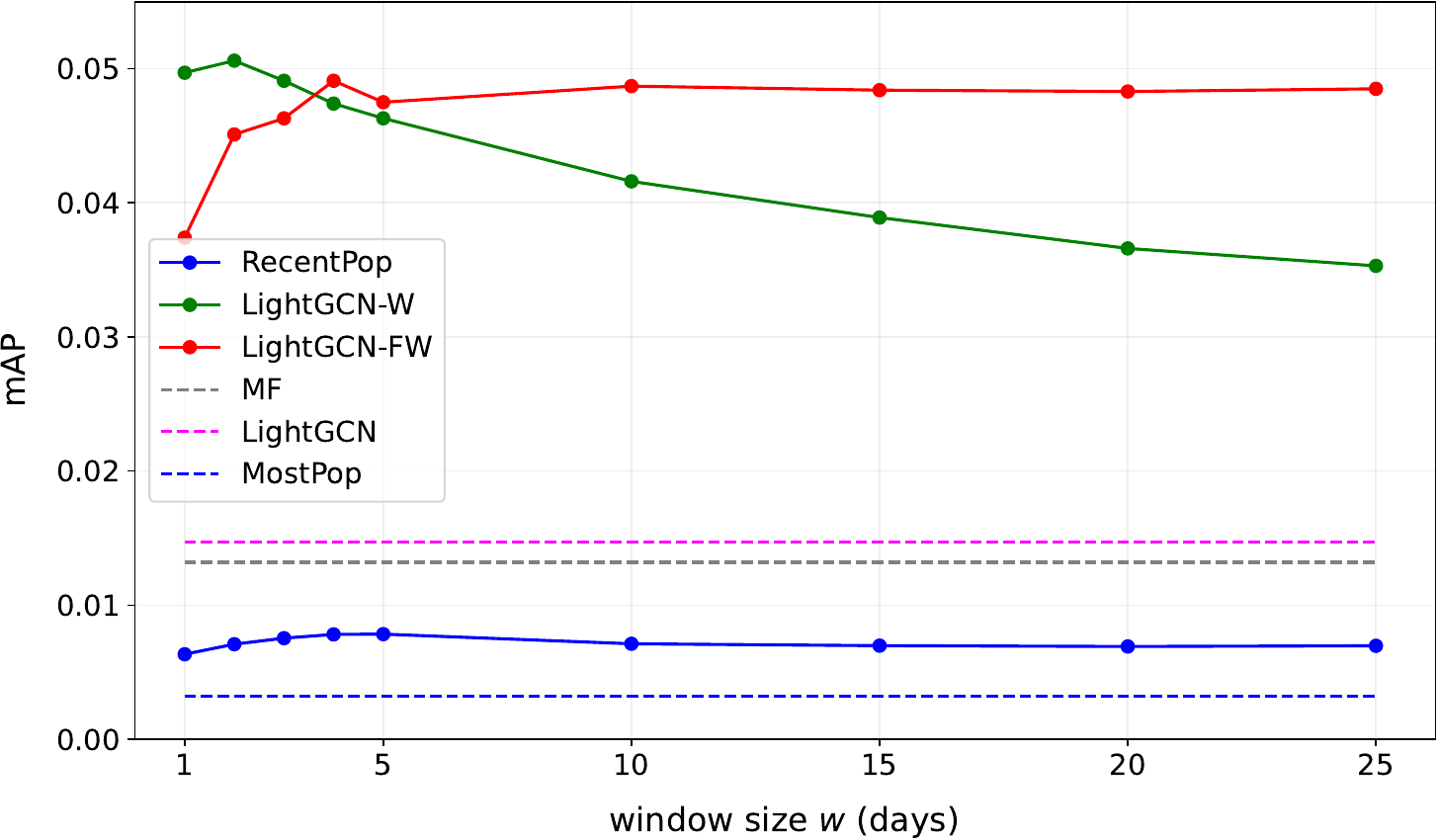}
\caption{\small Average model performance for various window sizes. Dashed lines denote a conventional algorithm trained using the full dataset. Only 1 layer and ID embeddings are used for GNN models.}
\label{fig:modelperformance}
\end{figure}

Figure~\ref{fig:modelperformance} illustrates the performance of the models discussed in Section~\ref{section:model}. Our models are trained on window sizes ranging from 1 to 25 days. We first notice that the best-performing model, LightGCN-W with $w=2$, significantly outperforms the standard LightGCN, increasing mAP by 244\% from $1.47$ to $5.06$. However, its performance declines as the window size increases, demonstrating that using more data does not necessarily lead to better performance. In fact, when using the entirety of the training data, LightGCN yields only a marginal improvement, enhancing mAP by 10.6\% over MF. 

The other model, LightGCN-FW, which assigns more weights to recent data, shows less sensitivity to the window size. Its performance seems to stabilize after $w=5$, even as older data is incorporated. Interestingly, even a straightforward baseline like MostPop shows improved performance as older data is discarded over time, highlighting the rapid shifts in the popularity of financial products. Notably, updating the popularity ranking of items daily using only the interactions of the previous $5$ days instead of the full history increases mAP from 0.32 (MostPop) to 0.78 (RecentPop).

These results highlight the effectiveness of our window-based approach in capturing dynamic user interests, emphasizing the importance of selecting the right window size for optimal performance in dynamic environments and prioritizing recent data to improve the accuracy of graph-based recommender systems.

\begin{table*}[!h]
\centering
\caption{\small Performance comparison between LightGCN and LightGCN-W for a varying number of GNN layers (rows) and different feature initialization modes (columns). The improvement in bold refers to the relative performance gain between the best model ($w=2$) and LightGCN. }
\label{tab:model-evaluation}
\begin{tabular}{
  @{}l 
  l 
  S[table-format=2.2]
  S[table-format=2.2]
  S[table-format=2.2]
  S[table-format=2.2]
  S[table-format=2.2]
  S[table-format=2.2]
  S[table-format=2.2]
  S[table-format=2.2]@{}
}
\toprule
& & \multicolumn{4}{c}{\textbf{ID}} & \multicolumn{4}{c}{\textbf{FEATS}} \\
\cmidrule(lr){3-6} \cmidrule(lr){7-10}
& & {mAP} & {MRR} & {NDCG@50} & {Recall@50} & {mAP} & {MRR} & {NDCG@50} & {Recall@50} \\
\midrule
\multirow{4}{*}{\textbf{1L}} & LightGCN & 1.47 & 3.78 & 2.98 & 6.50 & 1.45 & 3.86 & 3.08 & 6.98 \\
                             & LightGCN-W (w=1) & 4.97 & 9.67 & 8.43 & 15.83 & 5.55 & 10.27 & 8.66 & 15.60 \\
                             & LightGCN-W (w=2) & 5.06 & 9.78 & 8.67 & 16.45 & 5.76 & 10.39 & 9.03 & 15.90 \\
                             & LightGCN-W (w=5) & 4.63 & 9.23 & 8.16 & 15.95 & 5.10 & 9.39 & 8.30 & 15.32 \\
                             & \textbf{Improvement} & \textbf{244\%} & \textbf{158\%} & \textbf{191\%} & \textbf{154\%} & \textbf{297\%} & \textbf{169\%} & \textbf{193\%} & \textbf{127\%} \\
\midrule
\multirow{4}{*}{\textbf{2L}} & LightGCN & 1.40 & 3.68 & 2.85 & 6.21 & 1.41 & 3.69 & 2.84 & 6.19 \\
                             & LightGCN-W (w=1) & 5.25 & 10.16 & 8.84 & 16.38 & 6.38 & 11.49 & 9.71 & 16.36 \\
                             & LightGCN-W (w=2) & 5.40 & 10.31 & 9.15 & 17.13 & 6.55 & 11.73 & 10.09 & 17.23 \\
                             & LightGCN-W (w=5) & 5.06 & 9.96 & 8.82 & 17.00 & 5.94 & 10.83 & 9.47 & 17.00 \\
                             & \textbf{Improvement} & \textbf{285\%} & \textbf{180\%} & \textbf{221\%} & \textbf{175\%} & \textbf{364\%} & \textbf{217\%} & \textbf{255\%} & \textbf{178\%} \\
\midrule
\multirow{4}{*}{\textbf{3L}} & LightGCN & 1.24 & 3.31 & 2.52 & 5.53 & 1.37 & 3.61 & 2.80 & 6.14 \\
                             & LightGCN-W (w=1) & 5.83 & 10.98 & 9.58 & 17.25 & 6.93 & 12.37 & 10.49 & 17.35 \\
                             & LightGCN-W (w=2) & 6.03 & 11.26 & 9.98 & 18.17 & 6.90 & 12.20 & 10.50 & 17.64 \\
                             & LightGCN-W (w=5) & 5.40 & 10.48 & 9.29 & 17.65 & 6.14 & 11.08 & 9.75 & 17.36 \\
                             & \textbf{Improvement} & \textbf{386\%} & \textbf{240\%} & \textbf{296\%} & \textbf{228\%} & \textbf{403\%} & \textbf{238\%} & \textbf{275\%} & \textbf{187\%} \\
\bottomrule
\end{tabular}
\end{table*}

\subsection{Ablation Study}

Table~\ref{tab:model-evaluation} compares LightGCN with our newly introduced method under different settings. Namely, we try three different layer sizes and two different initialization modes. We limit the comparison to the main variant, LightGCN-W, as it has a similar architecture to LightGCN. Also, we only present the three window sizes, $w \in [1, 2, 5]$. Our method consistently outperforms LightGCN across all configurations and in all metrics.

First, the performance improvements achieved by our model become more pronounced as the number of layers increases. For instance, with 3 layers, our model achieves the highest percentage improvements compared to the 1-layer and 2-layer models. Conversely, LightGCN's performance decreases with additional layers. This suggests that our approach is more effective at capturing high-order collaborative signals in deeper models, whereas LightGCN does not benefit similarly from increased layer depth.

Second, using additional categorical features enhances the performance of both approaches in most cases. This indicates that adding features enables the models to learn richer representations, thus leading to better performance. Importantly, this enhancement is more significant in our model, as reflected in the greater percentage improvements when features are used compared to LightGCN.

The performance degradation of LightGCN with added layers likely stems from its reliance on the entirety of transaction history, which can only capture a global interest at prediction time. For example, consider user $u$ who made numerous purchases of item $i_1$ over the past years but has recently bought item $i_2$ once. Even if her interest has shifted or item $i_1$ has become unavailable, her embedding will continue to be dominated by $i_1$, overshadowing recent emerging preferences. Adding more layers aggravates this issue by amplifying the influence of older transactions as they will be used to propagate outdated collaborative signals, reducing the model's responsiveness to current preferences.
 
In contrast, our approach ensures the graph represents only recent user preferences. We hypothesize that in dynamic domains like finance, users who interact with the same items within a shorter time period show stronger similarities. Hence, our approach enables the model to capture high-order collaborative signals from the most relevant interactions, leading to improved performance.
 



\section{Conclusion}

In this work, we present a novel extension for LightGCN designed to capture dynamic user interests using causal convolutions. Our findings highlight the importance of maintaining the order of interactions and prioritizing recent ones for delivering accurate recommendations in dynamic domains like finance, where past data quickly becomes irrelevant. We demonstrate that selecting the appropriate window size can significantly improve the model performance without necessitating architectural modifications. Further analysis reveals that our model exhibits a higher capacity for improvement with more layers and additional features compared to LightGCN. In future work, we plan to explore alternative temporal sampling mechanisms that maintain causality while ensuring a balanced representation across diverse client profiles. Additionally, we aim to extend the framework to make it more comprehensible by incorporating dynamic features like real-time market data and developing further components to model more complex behaviour. 


\bibliographystyle{ACM-Reference-Format}
\bibliography{references}

\appendix

\section{Appendix}

\subsection{Implementation Details}
\label{appendix:implementation_details}

The embedding sizes reported at the end of Section~\ref{section:training} correspond to those of the best model and were searched in $[16, 32, 64, 128]$ for the ID embeddings and in $[8, 16, 32, 64]$ for the embeddings of each categorical feature. The models have a relatively small number of parameters to tune. Apart from what was mentioned, we use a learning rate of 1e-4 and train for 40 epochs with early stopping, where patience is set to 10.

We implement our approach using PyTorch Geometric\footnote{\url{https://github.com/pyg-team/pytorch_geometric}}. For the data part, we use a list of PyG graphs to represent the dataset as a series of temporal snapshots. As for the model part, we rely on the implementation of LGConv layer provided by the framework and we modify it for our needs.

\subsection{Pseudo-code}
\label{appendix:algorithm}

We provide the following pseudo-code to facilitate the implementation of our work, focusing solely on ID embeddings for simplicity.

\begin{algorithm}[!h]
\SetAlgoLined
\SetKwInOut{Input}{Input}
\SetKwInOut{Output}{Output}
\Input{List of graphs [$G_1, G_2, \ldots, G_T$], \\ Number of days $T$, \\ Window size $w$}
\Output{Dynamic user embeddings $\pmb{h}_u^{1:T}$, \\ Dynamic item embeddings $\pmb{h}_i^{1:T}$}

Initialize trainable node embeddings $\pmb{h}_u^0$ and $\pmb{h}_i^0$\;

\For{each day $t$ from $1$ to $T$}{
    Construct $G_{[t-w,t)}$ using transactions within $[t-w, t)$\;
        \For{each layer $k$ from $1$ to $K$}{
            \For{each node $u \in G_{t}$}{
                $\pmb{h}_u^k(t) \leftarrow \sum_{i \in \mathcal{N}_{t,w}(u)} c_{ui}^{\Delta t} \cdot \pmb{h}_i^{k-1}(t)$\;
            }
            \For{each node $i \in G_{t}$}{
                $\pmb{h}_i^k(t) \leftarrow \sum_{u \in \mathcal{N}_{t,w}(i)} c_{ui}^{\Delta t} \cdot \pmb{h}_u^{k-1}(t)$\;
            }
    }
    // Aggregate layer embeddings (eq.~\ref{eq:layer_aggregation}):\\
        $\pmb{e}_u^t \leftarrow \sum_{k=0}^{K} \alpha_k \cdot \pmb{h}_u^k(t)$ \\ 
        $\pmb{e}_i^t \leftarrow \sum_{k=0}^{K} \alpha_k \cdot \pmb{h}_i^k(t)$\;
        
    Sample negative triplets for BPR\;
    Compute $\mathcal{L}_{BPR}$ (eq.~\ref{eq:bpr_loss})\;
    Back-propagate and update model parameters\;
}
\Return user and item embeddings: $\pmb{h}_u^{1:T}$, $\pmb{h}_i^{1:T}$\;
\caption{Causal Graph Convolution}
\end{algorithm}

\begin{figure*}[!b]
\centering
\includegraphics[width=0.75\textwidth]{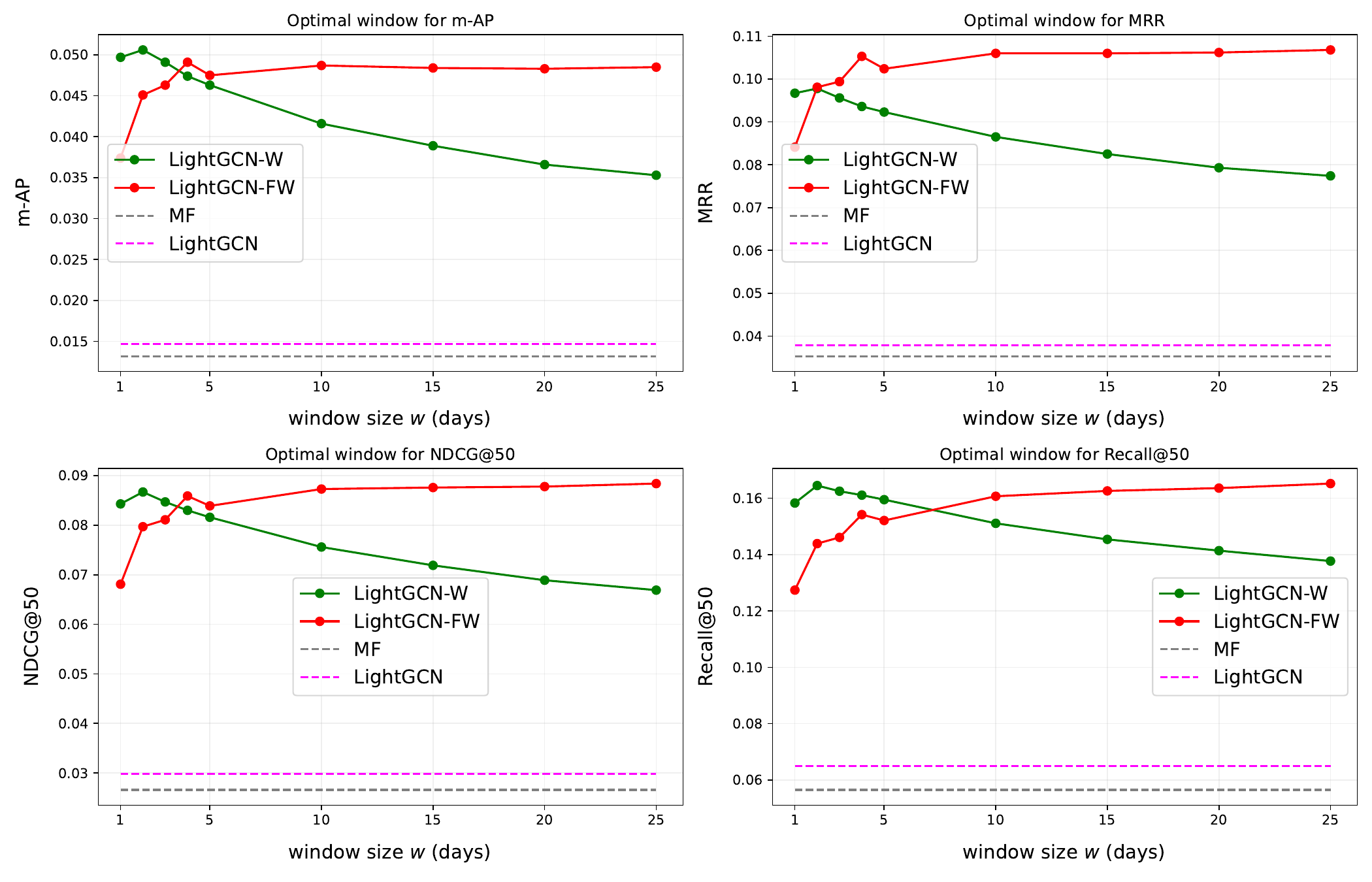}
\caption{\small Performance comparison of the main models across different metrics with varying window sizes.}
\label{fig:full_analysis}
\end{figure*}

\subsection{Ranking Metrics}
\label{appendix:detailed_metrics}

This section provides a more technical definition of the ranking metrics used in Section~\ref{section:eval_metrics}, introducing the necessary notations and formulas for each metric.

Let $Q_t = \{\,u \mid (u,i,t) \in \mathcal{E}_t \, \}$ be the set of clients who have at least one positive interaction on day $t$. For each client $u$, the model generates a list of recommended items $R_u^t = [i_1, \dots, i_L]$, where $L$ represents the number of available items on day $t$. The items are sorted by decreasing order of the predicted scores $\hat{a}_{ui}^t$. 

We define the indicator function $rel_u^t(j)$ to be to 1 if the item at rank $j$ is relevant (i.e., the user has interacted with), 0 otherwise. 

Let $T_u^t = \sum_{j=1}^{L} rel_u^t(j)$ represent the total number of relavant items for user $u$ at time $t$.

Given the above notations, the metrics are defined as follows:

\begin{equation}
	\text{MRR}_t = \frac{1}{\mid Q_t \mid}  \sum_{u \sim Q_t} \frac{1}{\text{min}(\{ \, j \mid rel_u^t(j) = 1 \, \})},
\end{equation}

\begin{equation}
	\text{Recall}_t@K = \frac{1}{\mid Q_t \mid}  \sum_{u \sim Q_t} \frac{\sum_{j=1}^{K} rel_u^t(j)}{T_u^t},
\end{equation}

\begin{equation}
	\text{mAP}_t = \frac{1}{\mid Q_t \mid}  \sum_{u \sim Q_t} \sum_{l=1}^{L} \frac{\sum_{j=1}^{l} rel_u^t(j)}{l} \cdot rel_u^t(l),
\end{equation}

\begin{equation}
	\text{DCG}_t@K = \frac{1}{\mid Q_t \mid}  \sum_{u \sim Q_t} \sum_{j=1}^{K} \frac{rel_u^t(j)}{log_2(j+1)},
\end{equation}

\begin{equation}
	\text{NDCG}_t@K = \frac{\text{DCG}_t@K}{\text{IDCG}_t@K}, \newline
\end{equation}

where $\text{IDCG}_t@K$ is the ideal $\text{DCG}_t@K$, representing the score of an ideal ranking that places all the relevant items at the top.

\subsection{Additional Results}
\label{appendix:add_results}

To ascertain whether different metrics benefit from varying window sizes, we extend the analysis presented in Section~\ref{section:history_size} to include a more comprehensive comparison across all metrics. The results in Figure~\ref{fig:full_analysis} show a consistent pattern among the different metrics introduced in Section~\ref{section:eval_metrics}. Specifically, using a window size of 2 days generally yields optimal performance for LightGCN-W across all metrics. Furthermore, the relative ranking of the models remains relatively consistent regardless of the chosen criterion, suggesting that varying the window size yields uniform improvements across multiple evaluation criteria.

Moreover, we notice that LightGCN-W excels at smaller window sizes, making it ideal for scenarios where the most recent interactions are critical. Meanwhile, LightGCN-FW maintains high and stable performance for larger window sizes, making it a more robust choice when tuning the window size is infeasible.

\end{document}